\documentclass{elsart}
\usepackage{epsfig,graphicx, amssymb}
\begin{document}
\runauthor{G. Litak}

\begin{frontmatter}
\title{Homoclinic transition to chaos in the Ueda 
oscillator with  external forcing
}   

\author[Lublin1,Wien]{Grzegorz Litak\thanksref{E-mail}}
\author[Lublin2]{, Arkadiusz Syta}
\author[Lublin1]{, Marek Borowiec}

\address[Lublin1]{Department of Applied Mechanics, Technical University of
Lublin, 
Nadbystrzycka 36, PL-20-618 Lublin, Poland}
\address[Wien]{Institut f\"{u}r Mechanik und Mechatronik, Technische Universit\"{a}t Wien, 
 Wiedner 
Hauptstra$\beta$e
8 - 10 A-1040 Wien, Austria}
\address[Lublin2]{Department of Applied Mathematics, Technical University
of
Lublin,
Nadbystrzycka 36, PL-20-618 Lublin, Poland}

\thanks[E-mail]{Fax: +48-815250808; E-mail:
g.litak@pollub.pl (G. Litak)}

\begin{abstract}
We examine the Melnikov criterion for transition to chaos in case of a
single degree of freedom nonlinear oscillator with the Ueda well  
potential and an external periodic excitation term. Using effective 
Hamiltonian  we have examined homoclinic orbits and cross-sections of stable and 
unstable manifolds which gave the condition of transition to chaos through
a homoclinic bifurcation.

\end{abstract}

\begin{keyword}
 nonlinear dynamics, vibrations, bifurcation, chaos
\end{keyword}
%nonlinear dynamics, 05.45.a
%statistical physics and nonlinear dynamics, 05.10.a
%Mechanical vibrations, 46.40.f
%low-dimensional Chaos, 05.45.Ac
\end{frontmatter}
{\em PACS}:  05.45.a, 46.40.f, 05.10.a, 05.45.Ac

\section{Introduction}
The Melnikov method 
 has
become  a classical approach
for predicting chaotic bechaviour in 
presence of saddle points
on the basis of cross-sections of stable and unstable manifolds 
\cite{Melnikov1963,Guckenheimer1983,Wiggins1990}.
Usually, it is applied explicitly to 
systems which 
possess homoclinic orbits in multiple well potential like Duffing double well, 
or pendulum  systems \cite{Guckenheimer1983,Wiggins1990,Tyrkiel2005}, or to the single well 
systems with a smooth 
potential barrier against an unstable 
solution \cite{Szemplinska1995,Litak2006,Litak2007}. 
 By signaling global 
 homoclinic transition it builds  a  
 condition for creation of fractal boundary between attraction basins and  chaos 
appearance provided that a 
vibration amplitude is  large 
enough to reach this boundary.
This is the reason why this condition
can be easiest fulfilled in the region of 
nonlinear resonance.
Comparing to other approaches \cite{Szemplinska1993,Kapitaniak1993,Kapitaniak1991}, 
the above scenario is 
so clear and instructive that 
the main stream of research
has been focused on
smooth multi- well potentials where the basins of attractions belong to separate 
wells.

In the present paper we 
will adopt the Melnikov method to an effective system described by 
double solutions. We will also investigate a possible fractal smearing of the 
basins of their attractions.
Identifying  a saddle point we will find homoclinic orbits 
there and finally define the corresponding Melnikov
criterion.

%f1
\begin{figure}[htb]
\centerline{  
\epsfig{file=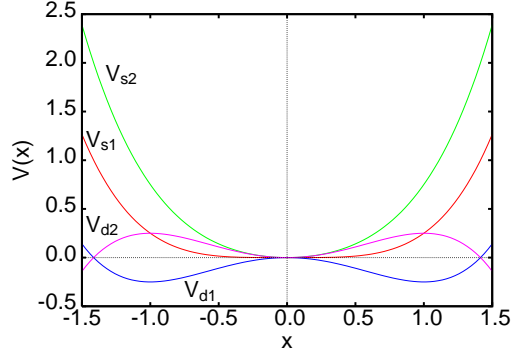,width=5.5cm,angle=-90}}
 \caption{ \label{fig1} 
Different possible potentials of a single  degree freedom system $V(x)=V_{s1}(x)$, 
$V_{s2}(x)$, $V_{d1}(x)$ and 
 $V_{d2}(x)$ as defined in Eq. \ref{eq2} for $\gamma=1.0$ and $\delta=1.0$.
}
\end{figure}

We start our analysis with one of the best known examples exhibiting
chaotic solution, namely, the
Ueda single well
system
%eq1
\begin{equation}
\label{eq1}
\frac{{\rm d}^2 x}{{\rm d}t^2} + \alpha  \frac{{\rm d} x}{{\rm
d}t}
+\gamma x^3=\mu \sin{ \Omega t},
\end{equation}
where $x$ is displacement  $\alpha \dot{x}$ is linear damping,
$\mu \sin{\Omega t}$ is an external excitation while
$-\gamma x^3$
is a cubic restore force ($\gamma > 0$).

%f2
\begin{figure}[htb]
\centerline{
\epsfig{file=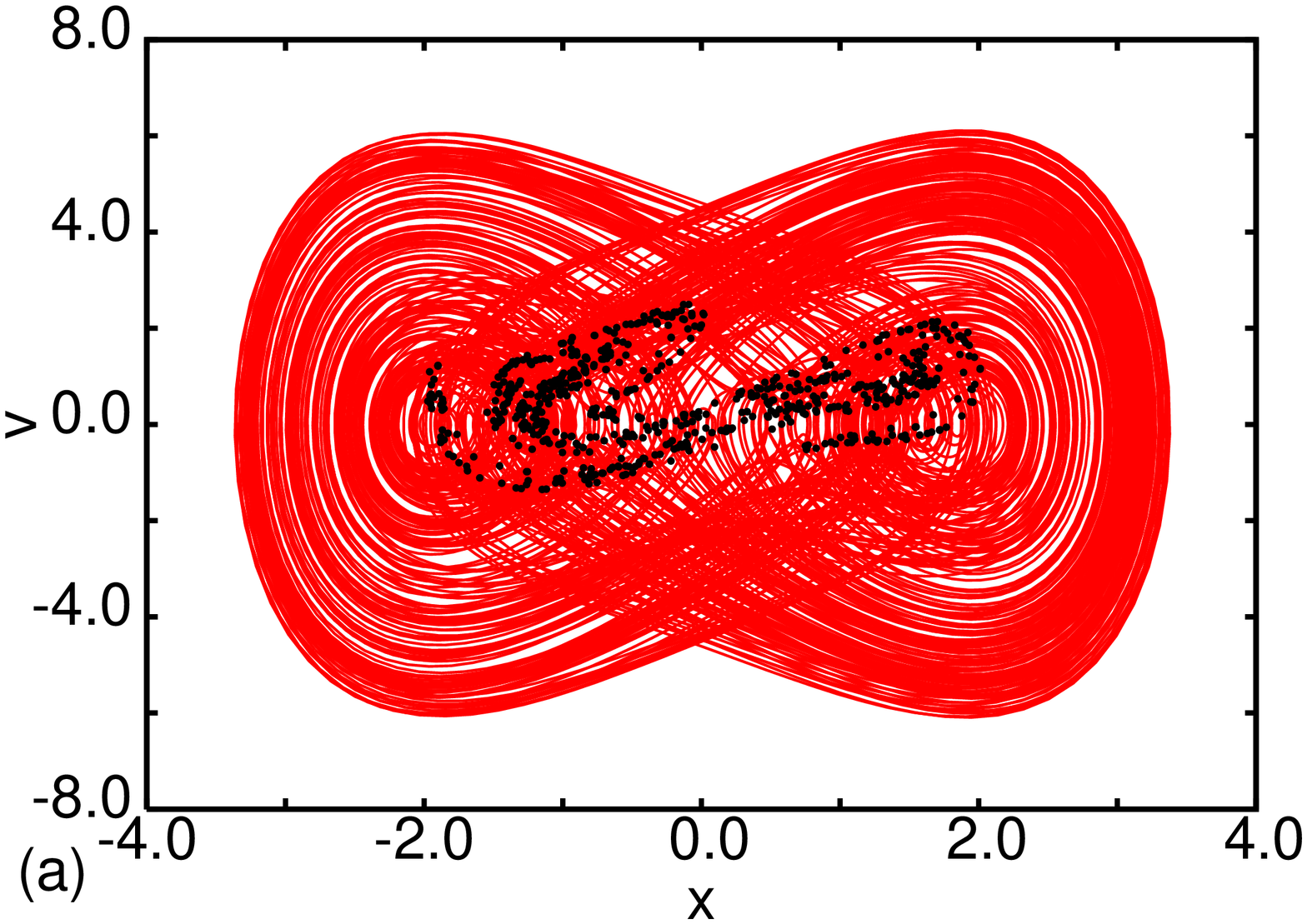,width=5.5cm,angle=-90} \hspace{-1cm} \epsfig{file=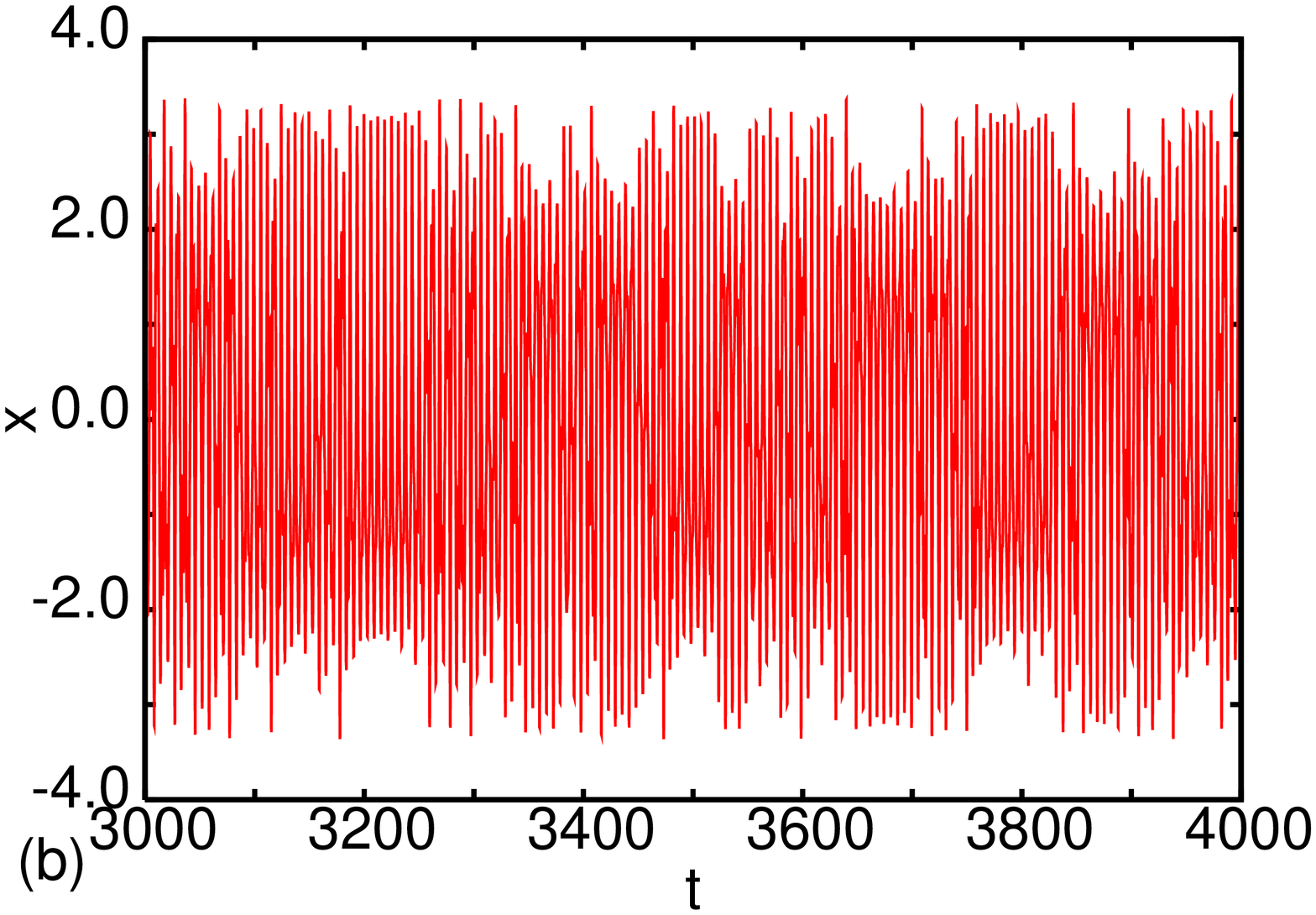,width=5.5cm,angle=-90}}
\centerline{
\epsfig{file=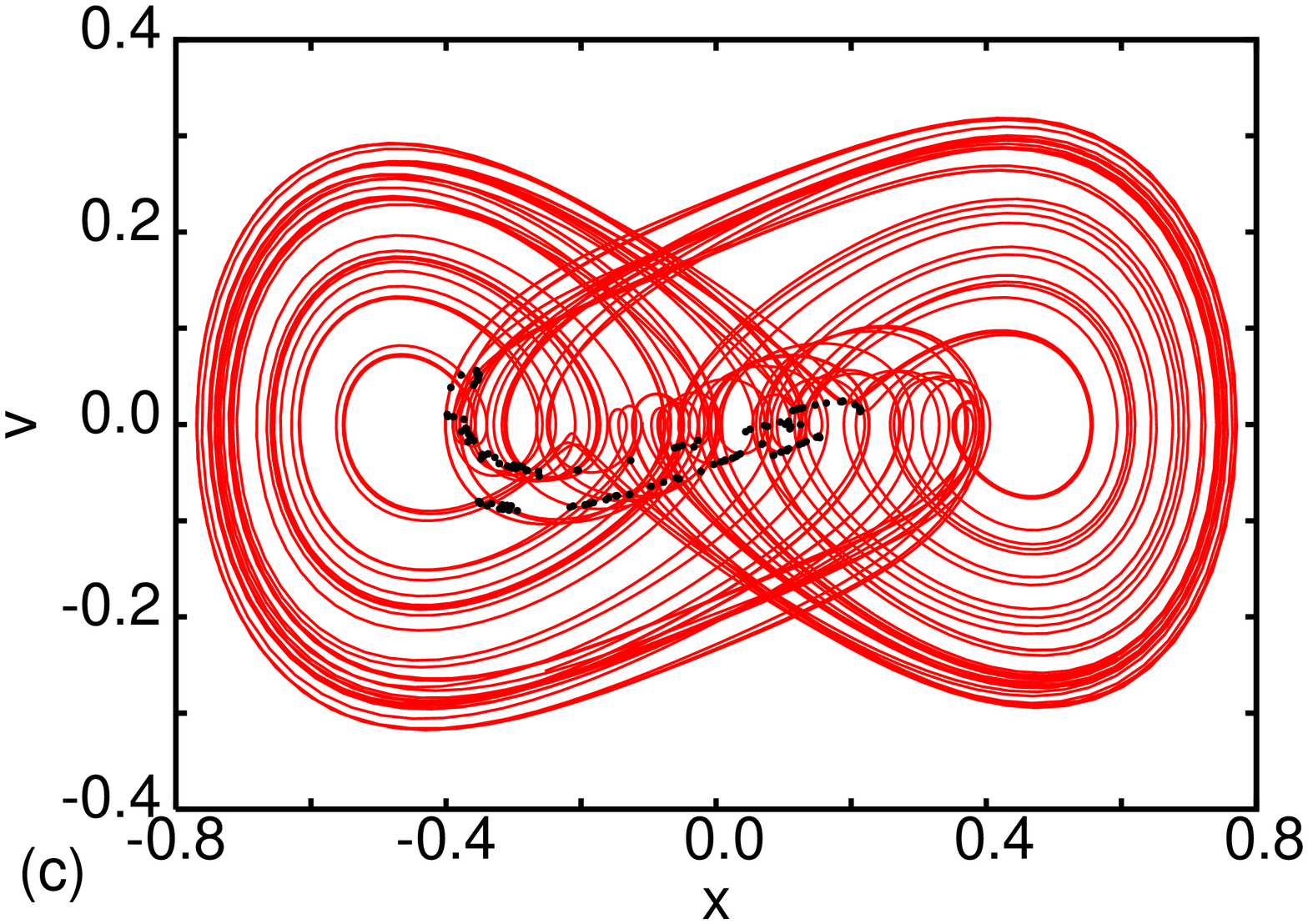,width=5.5cm,angle=-90} \hspace{-1cm} \epsfig{file=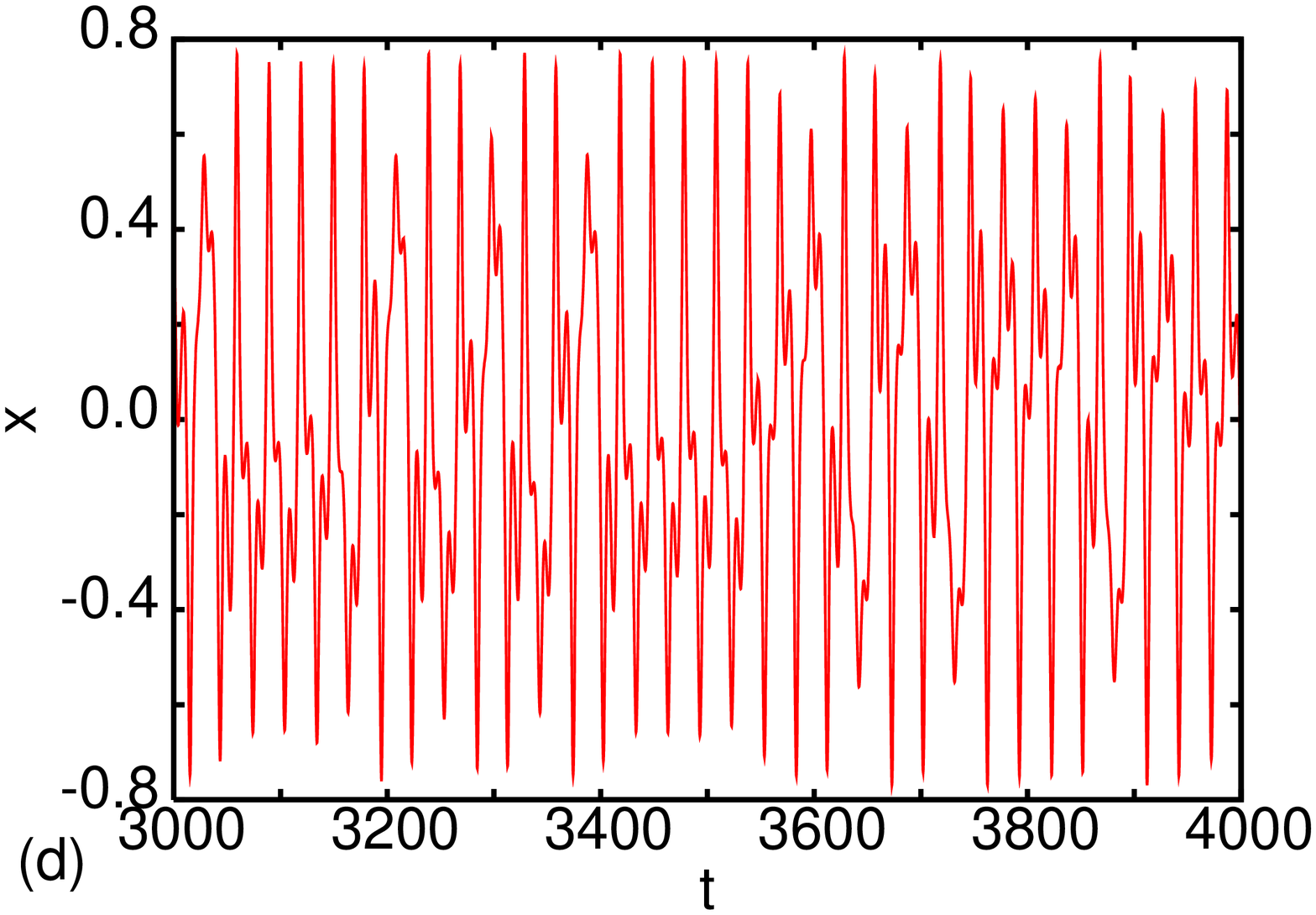,width=5.5cm,angle=-90}}
 \caption{ \label{fig2} Phase portraits (Fig. \ref{fig2}a,c - with lines) and Poincare maps
(Fig. \ref{fig2}a,c - with points) and corresponding time histories $x(t)$ (Fig. \ref{fig2}b,d) 
for two 
sets of 
system (Eq. \ref{eq1}) parameters:
Fig. \ref{fig2}a,b with $\gamma=1.0$, $\mu=7.5$, $\alpha=0.05$, $\Omega=1.0$,  Fig. 
\ref{fig2}c,d with 
$\gamma=1$
$\mu=0.1$,  $\alpha=0.05$,
$\Omega=0.21$. Top Lyapunov exponents for cases Fig. \ref{fig2}a,b and  Fig. \ref{fig2}c,d are
$\lambda_1=0.111$ and $\lambda_1=0.021$, respectively. The initial conditions used 
in both cases $x(t=0)=0.0$, $v(t=0)=2.5$.
}
\end{figure}

The above example is known from the pioneering Ueda work  on chaotic systems
\cite{Ueda1979}. Note that there are substantial differences between a single Ueda well potential 
$V_{s1}$ without a linear term and a double well Duffing potential $V_{d1}$ or an upside-down 
reflected double well 
potential
$V_{d2}=-V_{d1}$. On the other hand $V_{s1}$ resembles Duffing potential with hard stiffness 
$V_{s2}$ 
\begin{eqnarray}
V_{s1}(x) &=&\gamma \frac{x^4}{4},~~~~V_{d1}(x)=\gamma \frac{x^4}{4} -\delta  \frac{x^2}{2}, \label{eq2} \\
V_{d2}(x)&=&-\gamma \frac{x^4}{4} +\delta  \frac{x^2}{2},~~~~V_{s2}(x)=\gamma \frac{x^4}{4} +\delta  
\frac{x^2}{2},    
\nonumber
\end{eqnarray}  
where $\gamma$ and $\delta$ are defined positive.
In case of $V_{d1}(x)$ and $V_{d2}(x)$ potentials have well defined 
inflexion points and extrema which 
correspond to unstable saddle 
fixed points (Fig. \ref{fig1}) while  potentials $V_{s1}(x)$ and $V_{s2}(x)$ are fundamentally different 
with a single minimum at $x=0$ and without any  inflexion points. Note that 
previous applications of 
Melnikov theory
based on existence of multiple extrema of type  $V_{d1}(x)$ or $V_{d2}(x)$.   

On the other hand, Chakraborty,  in his recent paper  
\cite{Chakraborty2006}, noticed that
one can expect to apply the Melnikov theory even for a single well potential
 $V_{s2}$ (Eq. \ref{eq2}) 
after defining a new coordinate
system.
Motivated by this conjecture we are going to construct the Melnikov function and
derive a necessary condition to
system transition into a chaotic motion region.

\begin{figure}

\centerline{
\epsfig{file=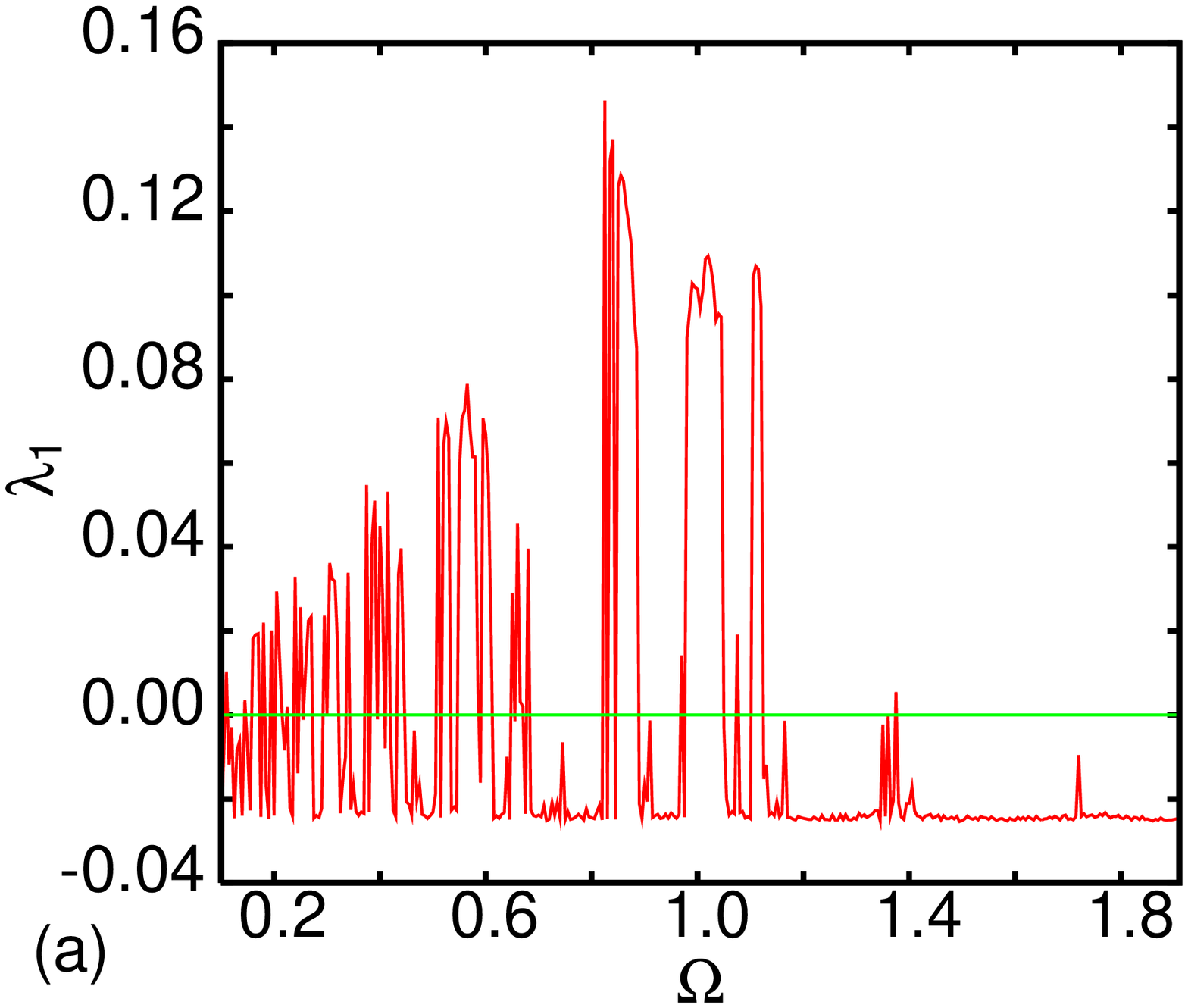,width=6.5cm,angle=-90}
\epsfig{file=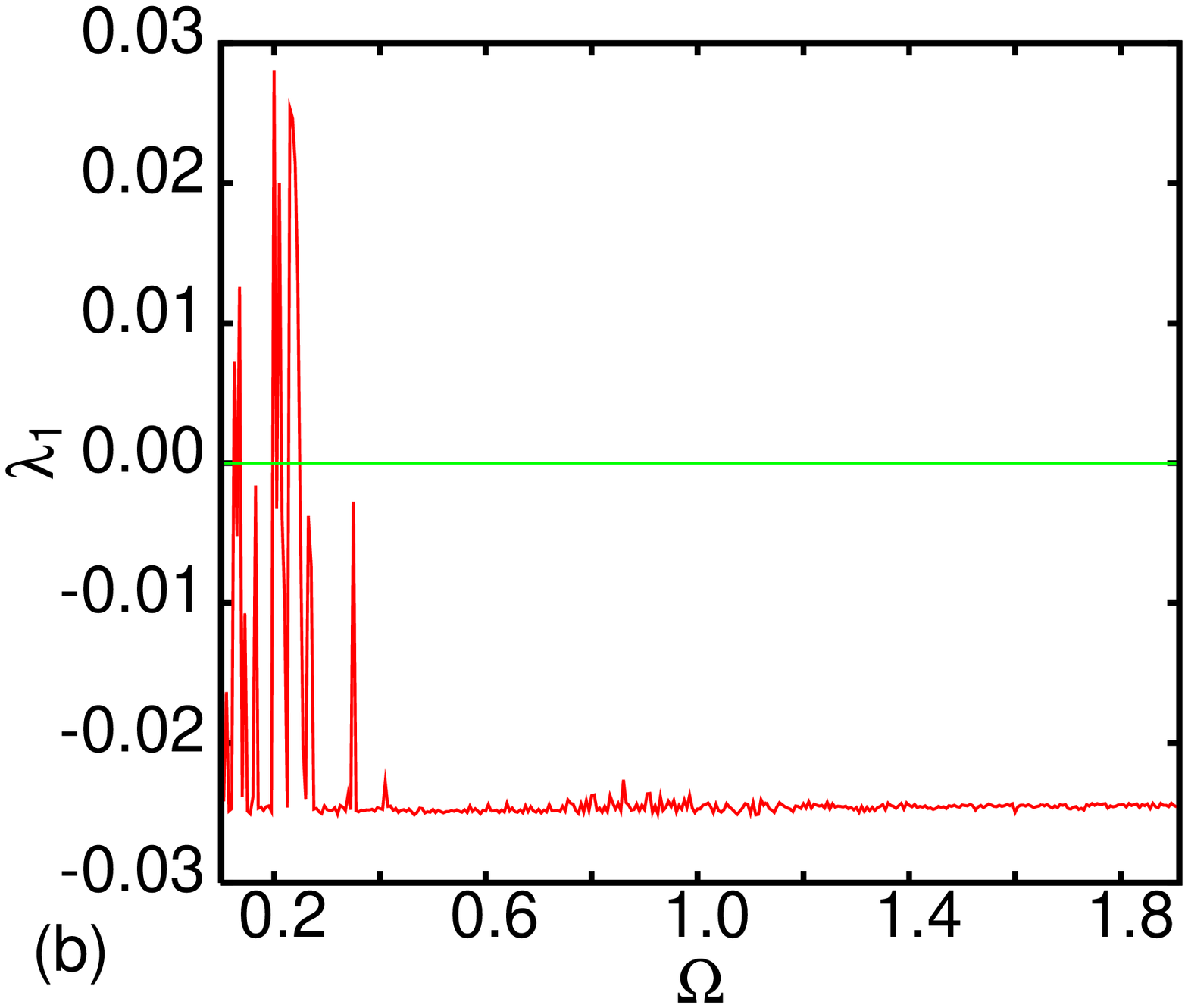,width=6.5cm,angle=-90}
}

\caption{ \label{fig3} Top Lyapunov exponents versus $\Omega \in [0.1:1.9] $ 
for two sets of system parameters;
$\gamma=1.0$, $\mu=7.5$, $\alpha=0.05$ (Fig  \ref{fig3}a) and Fig. with 
$\gamma=1.0$
$\mu=0.1$,  $\alpha=0.05$ (Fig. \ref{fig3}b). For the smallest $\Omega$ the initial 
condition we assumed to be $x(t=0)=0.0$, $v(t=0)=2.5$ while for any next larger $\Omega$
the final state values of $x$ and $v$ have been used as the initial conditions.
}
\end{figure}

To explore this possibility further we have performed numerical simulations to find the regions of 
chaotic solutions (Eq. \ref{eq1}).
In Fig. \ref{fig2} we show the phase portraits and Poincare maps as well as time histories of 
chaotic solutions. Note Figs. \ref{fig2}a-b correspond to the original Ueda system \cite{Ueda1979}, where
$\gamma = 1.0$, $\mu = 7.5$,  $\alpha = 0.05$,  $\Omega= 1.0$. Unfortunately,
the large value $\mu$ makes any perturbation method non-relevant  while
Note Figs. \ref{fig2}c-d, show the chaotic solution  for other choice of system parameters:  $\gamma 
= 1.0$, $\mu = 0.1$,  
$\alpha = 0.05$,  
$\Omega= 
0.21$. 
These chaotic solutions are characterized not only by the fractal strange attractors and non-periodic
time series (Figs. \ref{fig2}a-d) but also by positive top Lyapunov exponents $\lambda_1=0.111$ 
and 
$\lambda_1=0.021$
(for $\mu = 7.5$ and  $\mu = 0.1$ respectively).
In this second case (\ref{fig2}c-d)  a perturbation expansion in terms of $\mu$ could be 
performed.
In spite of different system parameters the attractors (given by Poincare maps Figs. \ref{fig2}a and
 \ref{fig2}c) look similar. 
To show the regions of chaotic solutions we have plotted the top Lyapunov exponent as a function of 
$\Omega$ in Fig. 
\ref{fig3}.
For larger $\mu$ we observe the relatively wide region of chaotic solutions (where the top Lyaounov 
exponent $\lambda_1$ has positive values). 
For smaller $\mu$ chaotic solutions are grouped in the region of low values of 
an excitation 
frequency $\Omega$ ($\Omega < 0.27$).  
Interestingly, the Lyapunov exponent, in that region $\lambda_1 \in [-0.03:0.03]$, is of the same 
ranges thus the change of $\mu$ 
does not scale them.   
Note this is a different region from that discussed by Chakraborty
\cite{Chakraborty2006} who concentrated on the region where the resonance curve possessed multiple 
solutions.  However in our particular system the chaotic solutions are evidently 
suppressed by higher excitation frequencies (Figs \ref{fig3}a-b).
Thus continuing our research on chaotic solutions appearance in the Ueda system (Eq. \ref{eq1})
(Figs. \ref{fig2}c-d and \ref{fig3}b)
 we will focus on low frequency region in this paper.

%3

Firstly we will look for homoclinic orbits
which can be treated analytically
by perturbation methods,
namely by the Melnikov method. Such a treatment has been
applied to selected problems in science and engineering
\cite{Guckenheimer1983,Wiggins1990}.

\section{Approximate solution around the main resonance}
In the vicinity of main resonance we assume periodic synchronized solution

%eq3
\begin{equation}
\label{eq3}
x=A(t)\sin(\Omega t) +B(t) \cos(\Omega t).
\end{equation}

introducing it to Eq. \ref{eq1}
and making use of the following trigonometric identities:
%eq4
\begin{eqnarray}
\label{eq4}
(\cos \psi)^3 &=& \frac{1}{4} \cos(3 \psi) + \frac{3}{4} \cos \psi, \nonumber \\
(\sin \psi)^3 &=&-\frac{1}{4} \sin(3 \psi) + \frac{3}{4} \sin \psi,
 \\
\sin(\psi)(\cos \psi)^2 &=& \sin(\psi)-\sin(\psi)^3=\frac{1}{4} \sin(3
\psi)
+ \frac{1}{4} \sin \psi, \nonumber \\
\cos(\psi)(\sin \psi)^2 &=& \cos(\psi)-\cos(\psi)^3=-\frac{1}{4} \cos(3
\psi)
+ \frac{1}{4} \cos \psi \nonumber
\end{eqnarray}

%f4
\begin{figure}

\centerline{
\epsfig{file=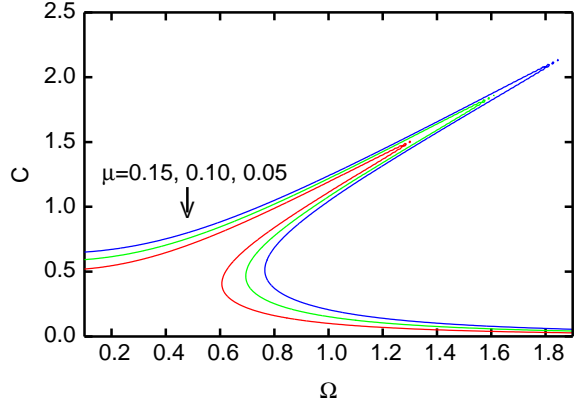,width=8.5cm,angle=-90}}

\vspace{-1cm}
~

\caption{ \label{fig4} Analytical amplitude $C$, calculated by harmonic balance approximation 
(Eq. \ref{eq7}), versus $\Omega \in [0.1:1.9]$ in the lowest order  for system 
parameters: 
$\gamma=1.0$
$\alpha=0.05$, and three values of $\mu$ ($\mu=0.05$, 0.10 and  0.15).}  
\end{figure}

we get
%eq5
\begin{eqnarray}
\label{eq5}
 \left[\frac{{\rm d}^2 A}{{\rm d}t^2} - \Omega^2 A + \frac{{\rm d}
A}{{\rm d}t} \alpha  -\alpha \Omega B
 +
\gamma \frac{3}{4} (A^3+AB^2) -\mu \right] \sin( \Omega t) \nonumber
\\
= \gamma (\frac{1}{4} A^3- \frac{3}{4}AB^2)
\sin(3 (\Omega
t))
\\
 \left[\frac{{\rm d}^2 B}{{\rm d}t^2} - \Omega^2 B + \frac{{\rm d}
B}{{\rm d}t} \alpha +\alpha \Omega A
 +
\gamma \frac{3}{4} (B^3+BA^2) \right] \cos( \Omega t) \nonumber
\\
= \gamma (-\frac{1}{4}B^3+\frac{3}{4}BA^2)
\cos(3 (\Omega
t)).
\nonumber
\end{eqnarray}

In the spirit of the harmonic balance approximation \cite{Kapitaniak1991} 
we find fixed
points
neglecting higher harmonics $\sin(3 \Omega t)$ and  $\cos(3 \Omega t)$  in 
the lowest
approximation.
Thus for $\dot A=0$ and  $\dot B=0$

%eq6
\begin{eqnarray}
\label{eq6}
- \Omega^2 A - \alpha \Omega B + \frac{3}{4} \gamma \left(A^3+AB^2 \right) =\mu, \\
- \Omega^2 B + \alpha \Omega A + \frac{3}{4} \gamma \left(B^3+BA^2 \right) = 0. \nonumber
\end{eqnarray}

after some simple algebra we get simple equation
%eq7
\begin{equation}
\label{eq7}
C^2 \alpha^2 \Omega^2 +
 C^2 \left( -\Omega^2 + \frac{3}{4} \gamma C^2 \right)^2 - \mu^2=0,
\end{equation}
where
%eq8
\begin{equation}
\label{eq8}
C^2=A^2+B^2.
\end{equation}

The result of an analytical solution (Eq. \ref{eq7}) for  the amplitude $C$ versus frequency
$\Omega$ has been shown in Fig. \ref{fig4}. One can see the characteristic incline in the 
resonance
in the right hand side. Above $\Omega=0.6$ there are triple solutions where the upper ad bottom
ones are stable and the middle one is unstable.

%f5
\begin{figure}

\centerline{
 \epsfig{file=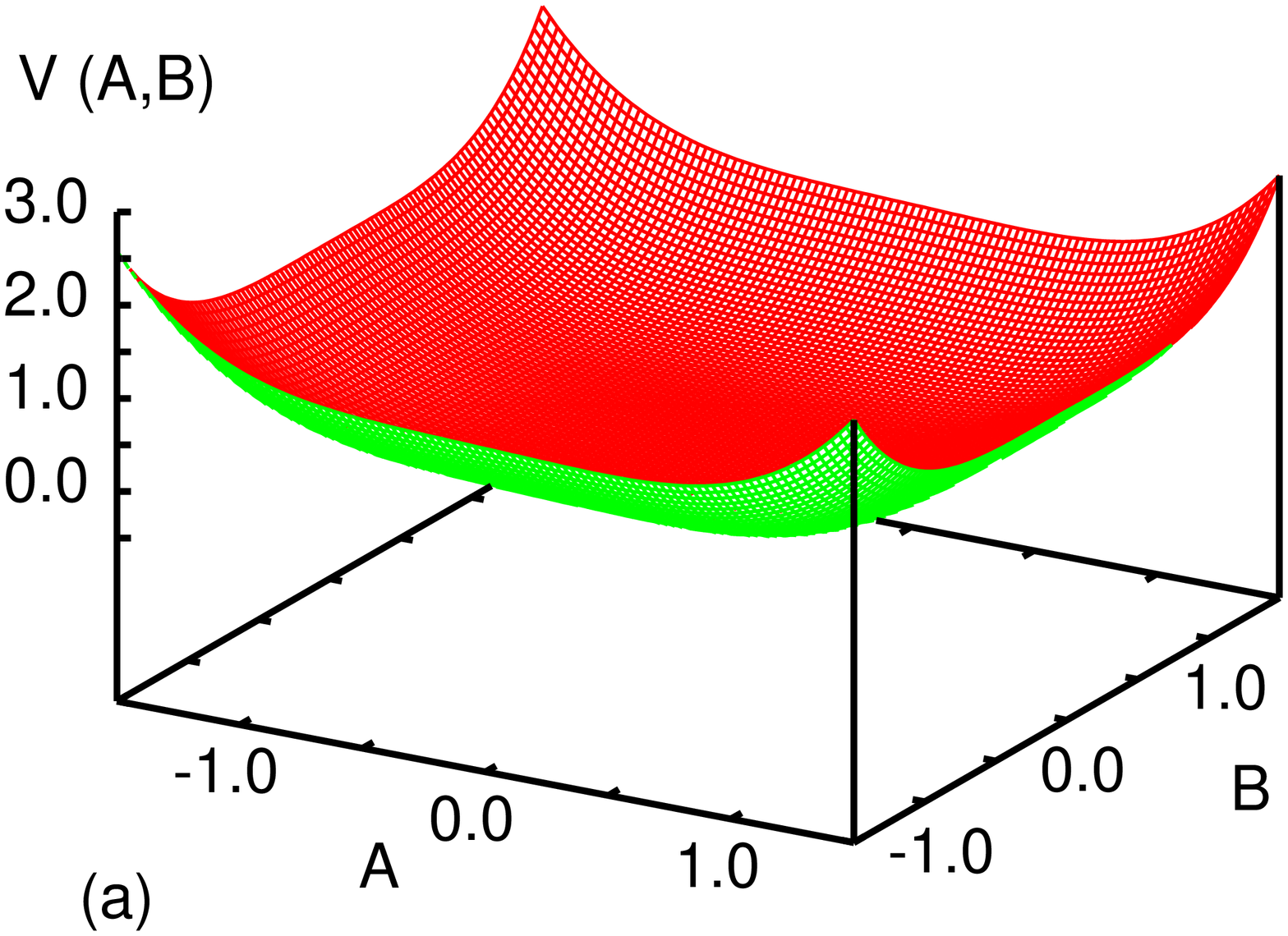,width=5.5cm,angle=-90} 
\epsfig{file=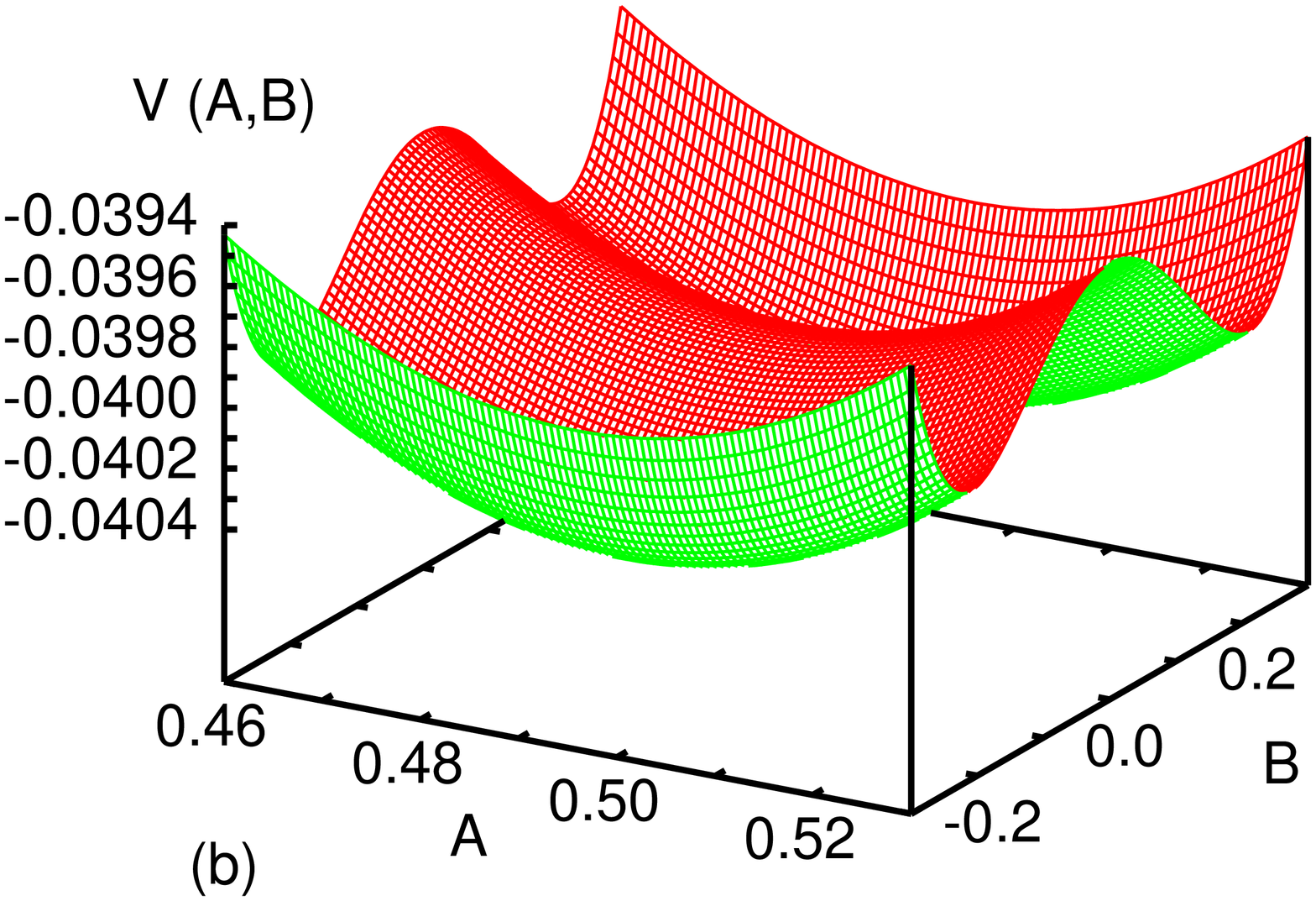,width=5.5cm,angle=-90}}

\caption{ \label{fig5}  Effective potential $V_{eff}= V_{eff}(A,B)$ for the following system 
parameters:  $\gamma=1.0$
$\mu=0.1$,  $\alpha=0.05$ and $\Omega=0.21$ plotted in two different scales (a) and (b) 
respectively.}
\end{figure}

\section{Melnikov approach beyond harmonic balance}

Motivated by Chakraborty \cite{Chakraborty2006} we now going beyond the harmonic balance approximation
keeping all terms of (Eq. \ref{eq4}) 
 Note, that the higher harmonic terms with $3 \Omega t$ can be easily transformed
into  $2 \Omega t$ and $\Omega t$ respectively
%eq9
\begin{eqnarray}
\label{eq9}
\cos(3 \Omega t) &=& \left(
2 \cos(2 \Omega t)-1\right) \cos(\Omega t), \\
\sin(3 \Omega t) &=& \left(
2 \cos(2 \Omega t)+1\right) \sin(\Omega t). \nonumber
\end{eqnarray}

Thus introducing Eqs. \ref{eq9} into Eq. \ref{eq5} and simplifying by
$\sin(\Omega t)$ and
$\cos(\Omega t)$
we get
%eq10
\begin{eqnarray}
\label{eq10}
\frac{{\rm d}^2 A}{{\rm d}t^2} &-& \Omega^2 A + \frac{{\rm d}
A}{{\rm d}t} \alpha  -\alpha \Omega B
 +
\gamma A^3 -\mu  \nonumber
=
\gamma \left(\frac{1}{2} A^3-\frac{3}{2} AB^2\right) \cos(2 \Omega
t), \\
 \frac{{\rm d}^2 B}{{\rm d}t^2} &-& \Omega^2 B + \frac{{\rm d}
B}{{\rm d}t} \alpha  +\alpha \Omega A
 +
\gamma B^3   =
\gamma \left(-\frac{1}{2} B^3-\frac{3}{2} BA^2\right)\cos(2 \Omega t). \nonumber \\ &~& 
\end{eqnarray}

In this way we have obtained new equations of motion for
new coordinates $A(t)$ and $B(t)$ with a parametric excitations.

Defining
velocities $v_A(t)$ and $v_B(t)$ and introducing small parameters into the equation $\epsilon$
and corresponding parameters $ \tilde \gamma$ and  $\tilde \alpha$ ($\epsilon \tilde \gamma = \gamma$,
$\epsilon \tilde \alpha = \alpha$)
 the first order equations of motion as
%eq11
\begin{eqnarray}
\frac{{\rm d} v_A}{{\rm d} t} &=& \Omega^2 A
- \gamma A^3  + \epsilon \left( -\tilde \alpha v_A  +\tilde \alpha B  +  \tilde \gamma
\left(\frac{1}{2} A^3-\frac{3}{2} AB^2\right)
\cos(2
\Omega
t) \right), \nonumber \\
\frac{{\rm d} A}{{\rm d} t} &= &v_A \label{eq11}
\end{eqnarray}
%eq12
\begin{eqnarray}
\frac{{\rm d} v_B}{{\rm d} t} &=& \Omega^2 B
- \gamma A^3  + \epsilon \left( -\tilde \alpha v_B - \tilde \alpha A  +  \tilde \gamma
\left(-\frac{1}{2} B^3-\frac{3}{2} A^2B\right)
\cos(2
\Omega
t) \right), \nonumber \\
\frac{{\rm d} B}{{\rm d} t} &= &v_B. \label{eq12}
\end{eqnarray}

The effective unperturbed Hamiltonian for the above set of equations (Eq. \ref{eq12}) can be written 
%eq13
\begin{equation}
H_{eff}=\frac{v_A^2}{2}+\frac{v_B^2}{2} +V_{eff}(A,B), \label{eq13}
\end{equation}
where 
%eq14
\begin{equation}
 V_{eff}(A)=\frac{-\Omega^2 A^2}{2}+ \frac{\gamma A^4}{4} -\mu A +\frac{-\Omega^2 B^2}{2}+ 
\frac{\gamma B^4}{4}, \label{eq14} 
\end{equation}
and $v_A$, $v_B$ are effective velocities which define kinetic terms of the Hamiltonian Eq. 
\ref{eq13}.
%f6
\begin{figure}

\centerline{
\epsfig{file=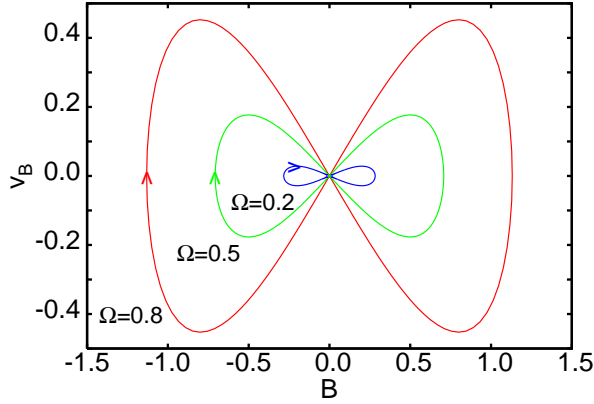,width=6.5cm,angle=-90}}

\caption{ \label{fig6}
Homoclinic orbits in ($B$, $V_B$) plane for $\Omega=0.8$, 0.5, 0.2, respectively}
\end{figure}

%f7
\begin{figure}

\centerline{
\epsfig{file=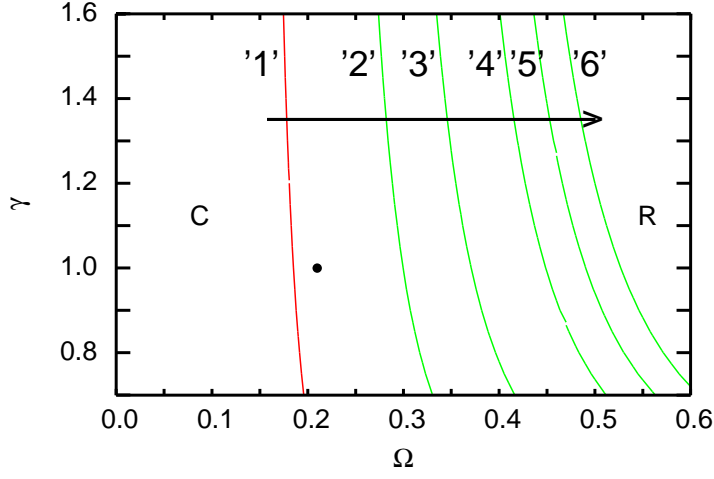,width=12.5cm,angle=-90}}
\vspace{-2.5cm}
~

\caption{ \label{fig7} Set of Melnikov curves plotted in a plane of
system ($\omega$,
$\gamma$). '1'--'6' corresponds to $\mu=0.01$, 0.03, 0.05, 0.08, 0.10 and 0.12,
respectively. C and R symbolize regions defined by the critical Melnikov curve
of possible
chaotic and  regular solutions.}
\end{figure}

%f8
\begin{figure}

\centerline{
\epsfig{file=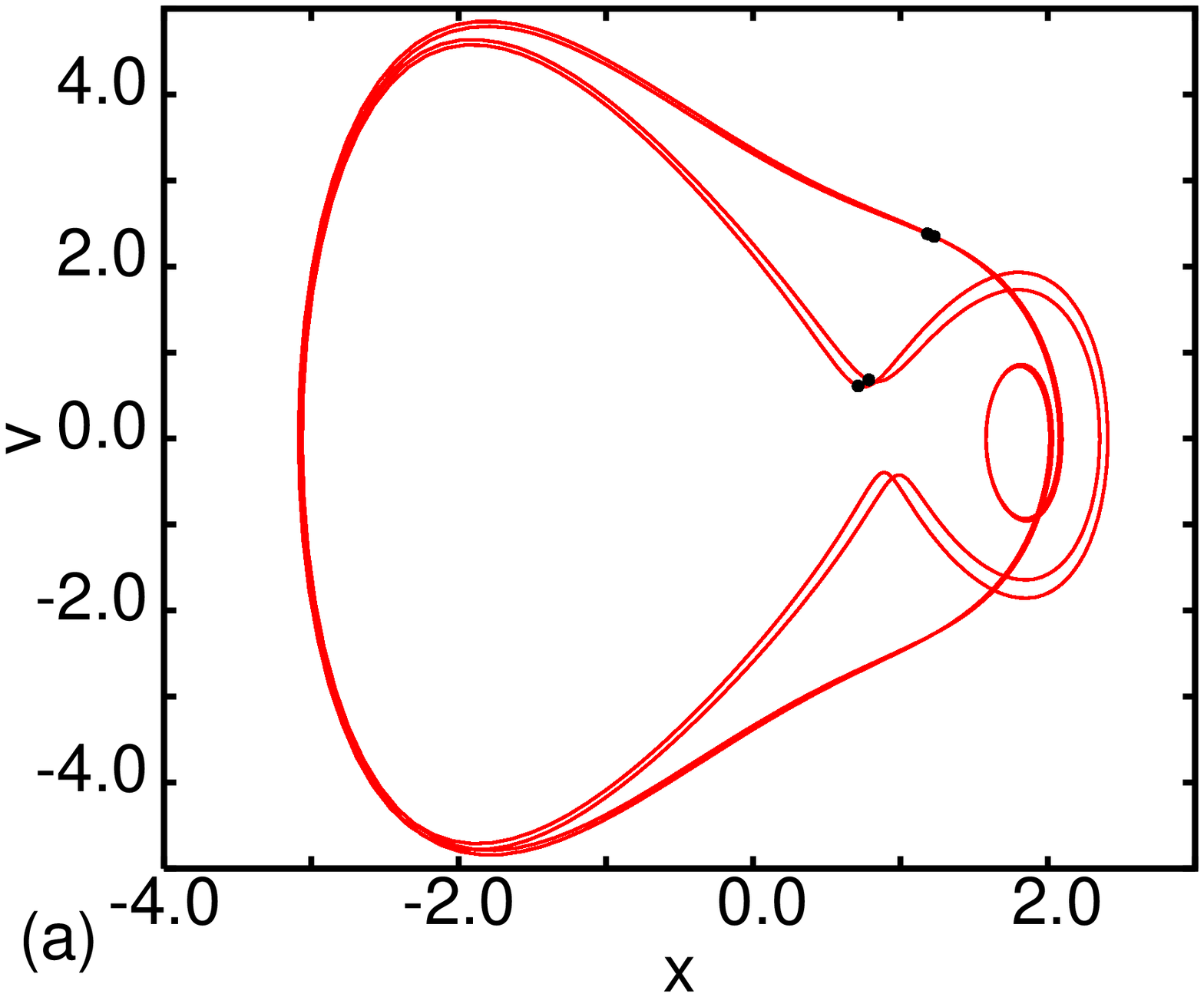,width=6.5cm,angle=-90} \hspace{-1cm}
\epsfig{file=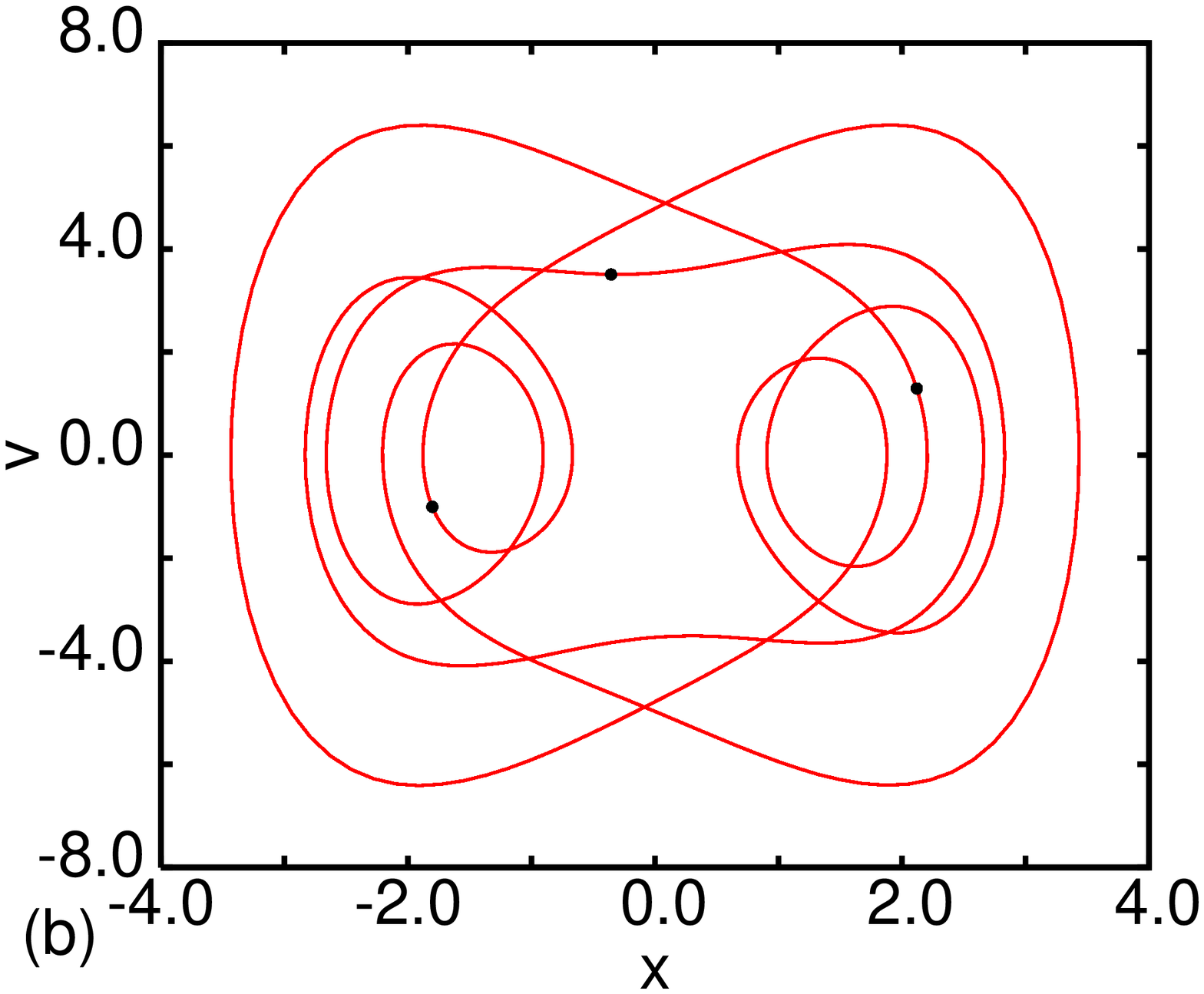,width=6.3cm,angle=-90}
}

\centerline{
\epsfig{file=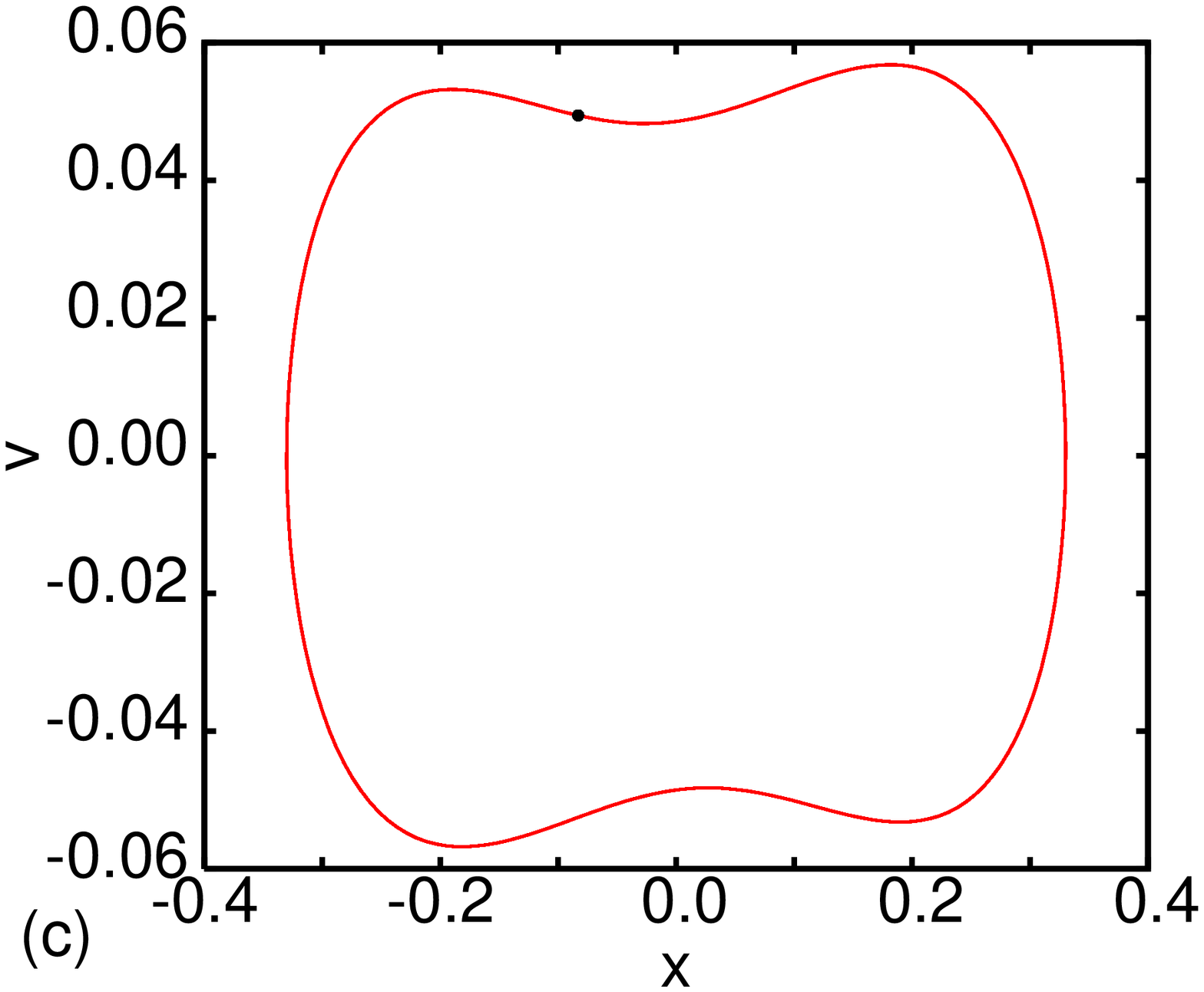,width=6.5cm,angle=-90} \hspace{-1cm}
\epsfig{file=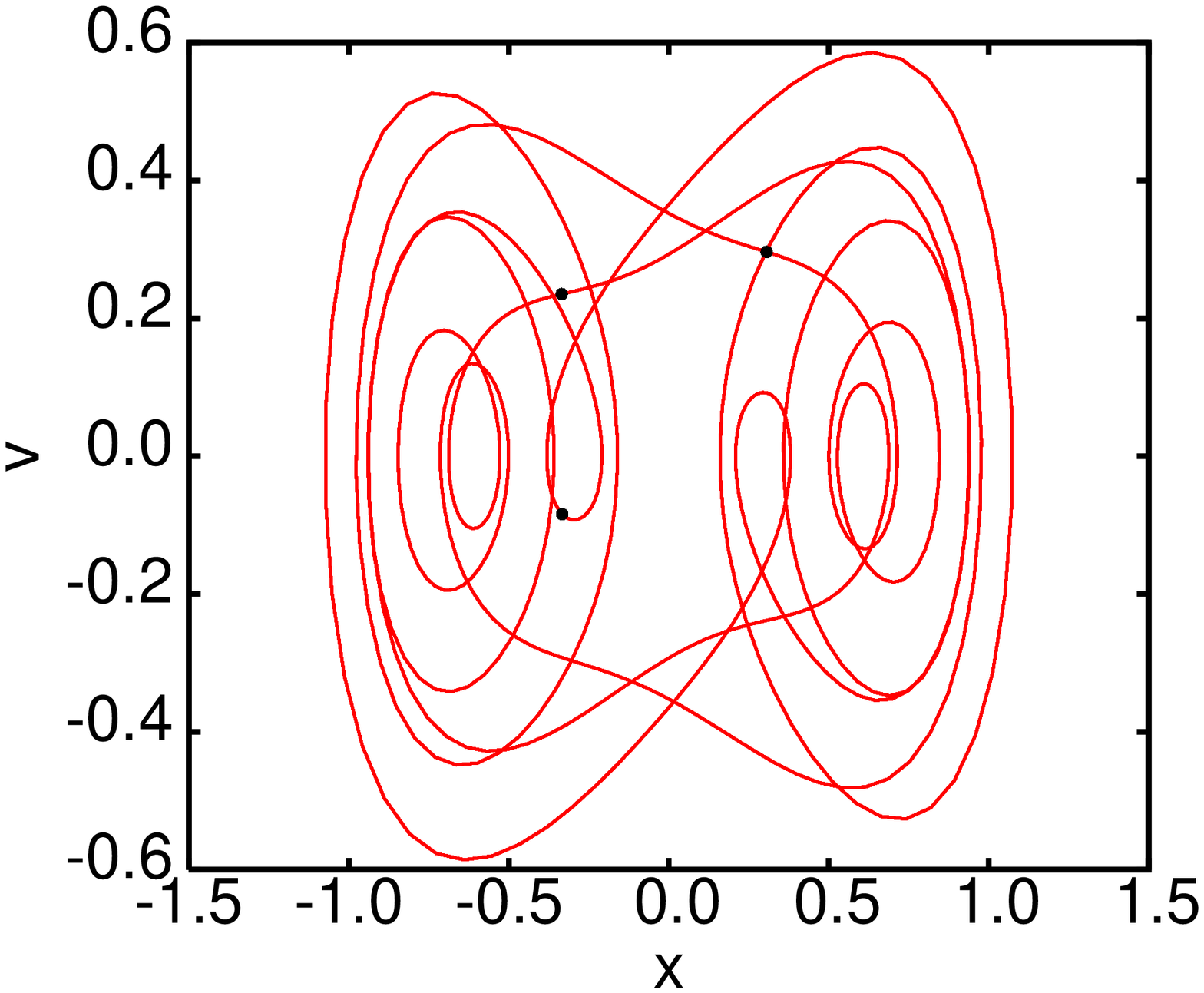,width=6.5cm,angle=-90}
}

\caption{ \label{fig8}  Phase portraits and Poincare maps for system parameters
chosen as follows:
$\Omega=1.1$, $\mu=7.0$ (Fig. 8a)
and $\mu=7.5$ (Fig. 8b);
$\Omega=0.21$, $\mu=0.01$ (Fig. 8c) and $\mu=0.35$ (Fig. 8d).
The initial condition used
in both cases $x(t=0)=0.0$, $v(t=0)=2.5$. The top Lyapunov exponent
$\lambda_1= 0.000$, -0.025, -0.024, -0.024 for Figs 8a-d respectively.
}

\end{figure}

This potential for the chosen system parameters,  $\gamma=1.0$
$\mu=0.1$,  $\alpha=0.05$ and $\Omega=0.21$, has been plotted  in Figs. \ref{fig5}a and b.
Note that for more precise mesh (Fig. \ref{fig5}b) 
we observe double-well structure of potential with degenerated minima  energy and 
a saddle point between them. 

Existence of this point with a horizontal tangent makes  
homoclinic bifurcations of the system possible i.e. transition from a regular to 
chaotic 
solution.

Note the characteristic saddle point $[A,B]=[A_0,0)$ is going to be reached in 
exactly defined albeit infinite time 
$t$  corresponding to $+\infty$ and $-\infty$
for stable and unstable orbits, respectively.
On the other hand $A_0$ can be obtained as the equilibrium fixed point from Eq. \ref{eq11}
and $\epsilon=0$
%eq15
\begin{equation}
A^3 - \frac{\Omega^2}{\gamma} A - \frac{\mu}{\gamma}=0. \label{eq15}
\end{equation}
Using the Cardano formula
%eq16
\begin{equation}
Q= -\frac{\Omega^6}{27\gamma^3} + \frac{\mu^2}{4\gamma^2}, ~~~~~~ 
A_0=\sqrt[3]{-\frac{\mu}{\gamma}+\sqrt{Q}}+
\sqrt[3]{-\frac{\mu}{\gamma} -\sqrt{Q}} \label{eq16}
\end{equation}

%f9
\begin{figure}

\centerline{
\epsfig{file=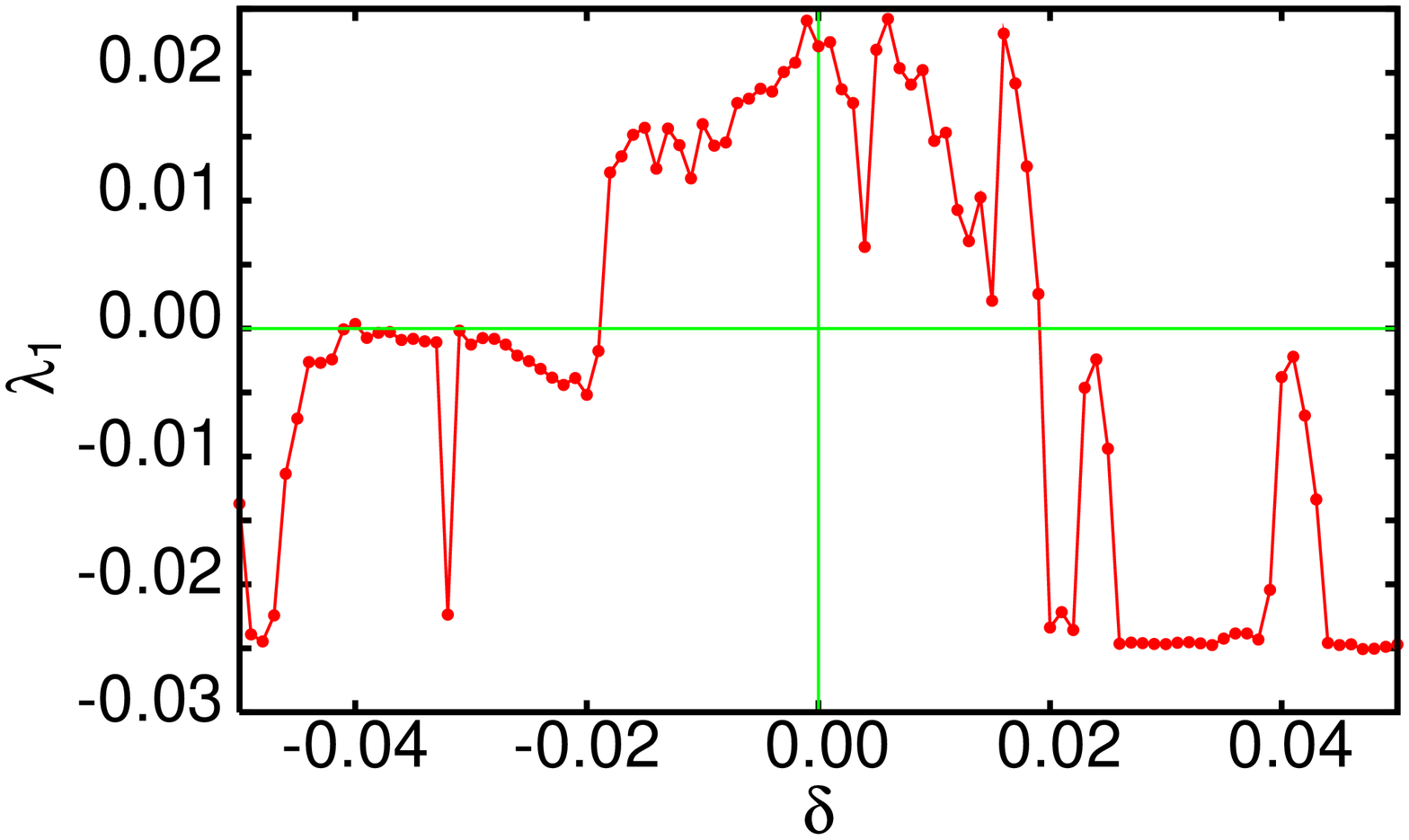,width=8.5cm,angle=-90}}

\vspace{-1cm}
~

\caption{ \label{fig9}
Top Lyapunov exponent against quadratic potential term $\delta$ as in $V_{s2}$ (Eq.
\ref{eq2}).
For the smallest (negative) $\delta$ the initial
conditions we assumed to be $x(t=0)=0.0$, $v(t=0)=2.5$ while for any next larger $\delta$
the final state values of $x$ and $v$ have been used as the initial conditions.
}
\end{figure}

The homoclinic orbit can be derived by assuming that 
%eq17
\begin{equation}
A=A_0 ~~~{\rm and}~~~ \lim_{t \rightarrow \pm \infty} B(t) =0. \label{eq17}
\end{equation}
In this case potential part for $B$ variable reads
%eq18
\begin{equation}
V_B'=- \frac{\Omega^2}{2}B^2 + \frac{\gamma}{4} B^4. \label{eq18}
\end{equation}

Assuming the condition $A=A_0$ and noting that $V_B=0$ at the saddle point ($B=0$) we perform 
standard analysis on energy 
conservation formula
%eq19
\begin{equation}
0=\frac{v_B^2}{2} + V_B'(B) \label{eq19}
\end{equation}
and after integration we get  
%eq20
\begin{equation}
- \frac{1}{\Omega} \ln{\left(\frac{\sqrt{\frac{2 \Omega^2}{\gamma}}+\sqrt{\frac{2 
\Omega^2}{\gamma}-B^2}
} {B} \right) } =t+t_0, \label{eq20}
\end{equation}
where $t_0$ is a time like integration constant.

Finally the homoclinic orbit is given by time dependent coordinate $B(t)=B^*(t)$
%eq21
\begin{equation}
B^*(t)= \sqrt{\frac{2}{\gamma}}\frac{\Omega}{\cosh{(\Omega (t+t_0))}} \label{eq21}
\end{equation}
and a corresponding velocity $v_B(t)=v_B^*(t)$
%eq22
\begin{equation}
v_B^*(t)= -\sqrt{\frac{2}{\gamma}}\frac{\Omega^2\sinh{(\Omega (t+t_0))}}{\cosh^2{(\Omega (t+t_0))}}.
\label{eq22}
\end{equation}
The homoclinic orbits for the same system parameters $\gamma$ and $\mu$ as in Fig. \ref{fig5} and assumed 
three values of excitation frequency  ($\Omega = 0.8$, 0.5, 0.2) has been plotted in Fig. \ref{fig6}.
Interestingly, for higher $\Omega$ we observe the effect of blowing the length of homoclinic loops 
and  
corresponding areas inside.

In case of perturbed orbits $W^S$ and $W^U$ the distance between them is given 
by 
the Melnikov function $M(t_0)$:
%eq23
\begin{equation}
M(t_0) = \int_{- \infty}^{ + \infty}  h( B^*, v_B^*)  \wedge g( B^*,
v_B^*) {\rm d} t \label{eq23}
\end{equation}
where the corresponding differential forms $h$ as the gradient of unperturbed 
Hamiltonian (Eq. \ref{eq13}) (for $A(t)=A_0$) leading to equations of motion 
%eq24
\begin{equation}
\frac{\partial H^0}{\partial B} = -\dot{v_B}, ~~~ \frac{\partial 
H^0}{\partial 
v_B} 
= \dot{B}, \label{eq24}
\end{equation}
while $g$ as its 
perturbation form of the above 
%eq25
\begin{eqnarray}
h &=& \left(\Omega^2 B + \gamma B^3\right) {\rm d} B  + v_B {\rm d}v_B, \label{eq25}\\
g &=& \left( -\tilde \alpha v_B - \tilde \alpha \Omega A_0  +  \tilde \gamma
\left(-\frac{1}{2} B^3-\frac{3}{2} BA^2 \right) \cos (2 \Omega t)  
\right)  {\rm d}x 
\nonumber
\end{eqnarray}
are
defined on homoclinic manifold $(B,v_B)=(B^*,v_B^*)$. 
Finally
the Melnikov integral is given by
%eq26
\begin{eqnarray}
M(t_0) &=&  \int_{- \infty}^{ + \infty} {\rm d}t~~ 
\left( -\tilde \alpha v_B^* - \tilde \alpha \Omega A_0  +  \tilde \gamma
\left(-\frac{1}{2} B^{*3}-\frac{3}{2} B^*A^{*2} \right) \cos (2 \Omega t)
\right) v_B^* \nonumber \\ ~~\label{eq26}
\end{eqnarray}
After substituting $B^*(t)$ and $v_B^*(t)$ by formulae given in Eq. 
\ref{eq6} and $A^*(t)=A_0$ (Eq. \ref{eq16})  
we get (see Appendix A) 

%eq27
\begin{eqnarray}
M(t_0)&=&-\frac{4}{3}\tilde{\alpha}\frac{\Omega^3}{\gamma}
+ \frac{16}{3}\tilde{\gamma}\frac{\Omega^4}{\gamma\sqrt{\gamma}}\pi {\rm e}^{-\pi}\sin(2\Omega t_0) \nonumber \\
&+& 12\tilde{\gamma}A_0^2\frac{\Omega^2}{\gamma} \pi {\rm e}^{-\pi}\sin(2\Omega t_0). \label{eq27}
\end{eqnarray}

In the Melnikov theory $M \sim d$ \cite{Melnikov1963,Guckenheimer1983,Wiggins1990}, where $d$ is the distance between 
stable and unstable manifolds.
Simple zero of the Melnikov function $M$ is associated with the cross-section of these manifolds 
indicating global homoclinic bifurcation.
In our case this condition (for the set of parameters: $\alpha$, $\gamma$, $\Omega$ and $A_0$ which a function of 
$\mu$, $\gamma$ and $\Omega$
 $A_0(\mu,\gamma,\Omega)$ (see Eqs.\ref{eq15}-\ref{eq16}) is fulfilled for $|\sin (2 \Omega 
t_0)|=0$
%eq28
\begin{equation}
\label{eq28}
 \alpha \Omega
 =(12\Omega^2 \sqrt{\gamma}  
 + 9 A_0^2 \sqrt{\gamma}) \pi {\rm e}^{-\pi}. 
\end{equation}
The analytical results for critical parameters $\gamma=\gamma_c$ and $\Omega=\Omega_c$ basing on this equation are 
shown in Fig. \ref{fig7} 
for  $\mu=0.01$, 0.03, 0.05, 0.08, 0.10 and 0.12. 
respectively. Left and right handed regions denoted for each curve as C and R respectively are 
related  to
possible "Chaotic" and "Regular" solutions. The results show that for larger $\gamma$ chaotic solution is more limited.
This tendency is better visible for a larger excitation amplitude $\mu$.
 Note the formerly investigated chaotic solution has been indicated by a singular point in the 
figure. Account for that 
case $\mu$ was chosen as 0.1 one can easily see that this solution match with the analytical 
prediction.
We have also checked that our system undergoes typical period doubling cascade showing also three 
points solution as well (defined on 
Poincare maps) Figs. \ref{fig8}b,d. In case of ) 
This regulars solution often accompany a chaotic solution.
Note also that for one of presented  solutions (in  Fig. \ref{fig8}a) the Lyapunov exponent 
was approximately 0
indicating a doubling period bifurcation point. A typical solution, synchronized with 
an excitation term, 
has been shown in Fig.  \ref{fig8}c.

\section{Summary and Conclusions}

In summary we have performed the Mielnikov analysis for the Ueda system with a single well potential. 
Through transforming the system to new variables it was possible to investigate the Mielnikov 
criterion for a global 
 homoclinic bifurcation from regular to chaotic oscillations. 

Our investigation was limited purposely to a small frequency $\Omega$ where the chaotic solutions emerge numerically
(Figs. \ref{fig2}c,d).  
However for different nonliner systems involving nonlinear damping terms and the self-excitation effects \cite{Litak1999}
the region of chaotic solutions could be different. The main simplification in our treatment was the assumption
$A(t)=A_0=$const.  In a more general  case one should expect additional time dependence of the amplitude $A$ which could create 
an additional 
shift
of the critical lines in Fig. \ref{fig7}. This shift should be dependent on $\Omega$, which  influences  strongly
the size of homoclinic orbits  Fig. \ref{fig6}.  
Consequently, for our system,  in the limit of large $\Omega$ the size of the homoclinc orbit is so large that condition 
$A(t)=$const. cannot be 
applied.

Our results  for the top Lyapunov exponent (Fig. \ref{fig9}) show also that the chaotic solution 
was preserved in
presence of a small linear force term $\delta$ (see $V_{s2}$ in Eq. \ref{eq2})
so the method presented here can be 
generalized to the Duffing system having 
linear and cubic force terms.

\section*{Acknowledgements}
%This research has been partially supported by the 6th Framework Programme,
%Marie Curie Actions, Transfer of Knowledge, Grant No. MTKD-CT-2004-014058.
This paper has been partially supported by the  Polish Ministry of Education.
GL would like to thank prof. H. Troger for helpful discussions.

\appendix{\Large \bf \noindent Appendix A}
\def\thesection{A}
\setcounter{equation}{0}
\def\theequation{A.\arabic{equation}}  % dla stylu 'article'

After substituting $B^*(t)$, $v_B^*(t)$ by formulae given in Eq.
\ref{eq6} and assuming that $A^*(t)=A_0$ as calculated in Eq. \ref{eq16} into Eq. \ref{eq23}, 
taking $\tau=\sqrt{\delta}t/2$ we get

%A1
\begin{eqnarray}
M(t_0) 
&=&-2\tilde{\alpha}\frac{\Omega^4}{\gamma}\int^{+\infty}_{-\infty}{\frac{\sinh^2( 
\Omega(t+t_0))}{\cosh^4(\Omega(t+t_0))}{\rm 
d}t} \nonumber \\
&+&
\sqrt{2}\tilde{\alpha}A_0\frac{\Omega^3}{\sqrt{\gamma}}\int^{ 
+\infty}_{-\infty}{\frac{\sinh(\Omega(t+t_0))}{\cosh^2(\Omega(t+t_0))}{\rm 
d}t} \label{A1}
\\
&+& 
2\tilde{\gamma}\frac{\Omega^5}{\gamma\sqrt{ 
\gamma}}\int^{+\infty}_{-\infty}{\frac{\sinh(\Omega(t+t_0))}{\cosh^5(\Omega(t+t_0))}\cos(2\Omega 
t){\rm 
d}t} \nonumber \\
&+& 
3\tilde{\gamma}A_0^2\frac{\Omega^3}{\gamma}\int^{+\infty}_{-\infty}{\frac{\sinh(\Omega(t+t_0))}{ 
\cosh^3(\Omega(t+t_0))}\cos(2\Omega 
t){\rm d}t}. \nonumber
\end{eqnarray}

Using $\tau = \Omega(t+t_0)$ we obtain
%A2
\begin{eqnarray}
 M(t_0)&=&-\tilde{\alpha}\frac{2}{\gamma}\Omega^3\int^{+\infty}_{-\infty}{\frac{\sinh^2{\tau}}{\cosh^4{\tau}}{\rm 
d}t} 
+
\tilde{\alpha}A_0\frac{\sqrt{2}}{\sqrt{\gamma}}\Omega^2\int^{+\infty}_{-\infty}{\frac{\sinh{\tau}}{\cosh^2{\tau}}{\rm 
d}t} \nonumber
\\
&+& 
\tilde{\gamma}\frac{2}{\gamma\sqrt{\gamma}}\Omega^4\int^{+\infty}_{-\infty}{\frac{\sinh{\tau}}{\cosh^5{\tau}}\cos(2\tau-2\Omega 
t_0){\rm 
d}t} \label{eqA2} \\
&+& 
\tilde{\gamma}A_0\frac{3}{\gamma}\Omega^2\int^{+\infty}_{ 
-\infty}{\frac{\sinh{\tau}}{\cosh^4{\tau}}\cos(2\tau-2\Omega 
t_0){\rm d}t}. \nonumber
\end{eqnarray}

The above Melnikov integral (Eq. \ref{eqA2}) can be written as 
%A3
\begin{eqnarray}
M(t_0)&=&-\tilde{\alpha}\frac{2}{\gamma}\Omega^3 I_1 + \tilde{\alpha}A\frac{\sqrt{2}}{\sqrt{\gamma}}\Omega^2 
I_2  \nonumber \\
&+& \tilde{\gamma}\frac{2}{\gamma\sqrt{\gamma}}\Omega^4 I_3 + 
\tilde{\gamma}A\frac{3}{\gamma}\Omega^2 I_4. \label{A3}
\end{eqnarray}

Integrals $I_1$ and $I_2$ can be calculated directly  
%A4
\begin{equation}
I_1 = \frac{2}{3}, \hspace{1cm} I_2 = 0.
\end{equation}	
Let us write $I_3$ and $I_4$ in the complex space as
%A5
\begin{eqnarray}
I_3&=&\cos(2\Omega t_0){\rm Re}\biggr\{\int^{+\infty}_{-\infty}{\frac{\sinh{\tau}}{\cosh^5{\tau}}\cos(2\tau){\rm 
d}t}+
i\int^{+\infty}_{-\infty}{\frac{\sinh{\tau}}{\cosh^5{\tau}}\sin(2\tau){\rm d}t}\biggl\} \nonumber
\\
&+&\sin(2\Omega t_0){\rm Im}\biggr\{\int^{+\infty}_{-\infty}{\frac{\sinh{\tau}}{\cosh^5{\tau}}\cos(2\tau){\rm d}t}+
i\int^{+\infty}_{-\infty}{\frac{\sinh{\tau}}{\cosh^5{\tau}}\sin(2\tau){\rm d}t}\biggl\},
\end{eqnarray}
and 
%A6
\begin{eqnarray}
I_4 &=&\cos(2\Omega t_0){\rm Re} \biggr\{\int^{+\infty}_{-\infty}{\frac{\sinh{\tau}}{\cosh^3{\tau}}\cos(2\tau)dt}+
i\int^{+\infty}_{-\infty}{\frac{\sinh{\tau}}{\cosh^3{\tau}}\sin(2\tau)dt}\biggl\} \nonumber \\
&+&\sin(2\Omega t_0){\rm Im} \biggr\{\int^{+\infty}_{-\infty}{\frac{\sinh{\tau}}{\cosh^3{\tau}}\cos(2\tau)dt}+
i\int^{+\infty}_{-\infty}{\frac{\sinh{\tau}}{\cosh^3{\tau}}\sin(2\tau)dt}\biggl\},
\end{eqnarray}
respectively.	

Now we can simplify the notation 
%A7,8
\begin{eqnarray}
I_3 &=&\cos(2\Omega 
t_0){\rm Re}\biggr\{\int^{+\infty}_{-\infty}{\frac{\sinh{\tau}}{\cosh^5{\tau}}{\rm e}^{2i\tau}dt}\biggl\} 
\nonumber \\
&+&\sin(2\Omega 
t_0){\rm Im}\biggr\{\int^{+\infty}_{-\infty}{\frac{\sinh{\tau}}{\cosh^5{\tau}}{\rm e}^{2i\tau}dt}\biggl\}, \\
I_4&=&\cos(2\Omega 
t_0){\rm Re}\biggr\{\int^{+\infty}_{-\infty}{\frac{\sinh{\tau}}{\cosh^3{\tau}}{\rm e}^{2i\tau}dt}\biggl\} \nonumber 
\\
&+&\sin(2\Omega 
t_0){\rm Im}\biggr\{\int^{+\infty}_{-\infty}{\frac{\sinh{\tau}}{\cosh^3{\tau}}{\rm e}^{2i\tau}dt}\biggl\}. 
\end{eqnarray}

Applying the residue theorem, we get 
%A9,10
\begin{eqnarray}
I_3&=&\cos(2\Omega t_0){\rm Re}\left(\frac{8}{3}\pi i(\cosh{\pi}-\sinh{\pi})\right) \nonumber \\
&+&\sin(2\Omega t_0){\rm Im}\left(\frac{8}{3}\pi 
i(\cosh{\pi}-\sinh{\pi})\right), \\
I_4&=&\cos(2\Omega t_0){\rm Re}(4\pi i(\cosh{\pi}-\sinh{\pi})) \nonumber \\
&+&\sin(2\Omega t_0){\rm Im}(4\pi 
i(\cosh{\pi}-\sinh{\pi})).
\end{eqnarray}

Consequently
%A11,12
\begin{eqnarray}
I_3 &=&\sin(2\Omega t_0)\frac{8}{3}\pi(\cosh{\pi}-\sinh{\pi}),
\\ I_4&=&\sin(2\Omega 
t_0)4\pi(\cosh{\pi}-\sinh{\pi}).
\end{eqnarray}

Finally the Melnikov integral reads 
%A13
\begin{eqnarray}
M(t_0)&=&-\frac{4}{3}\tilde{\alpha}\frac{\Omega^3}{\gamma}
+ \frac{16}{3}\tilde{\gamma}\frac{\Omega^4}{\gamma\sqrt{\gamma}}\pi {\rm e}^{-\pi}\sin(2\Omega t_0) \nonumber \\
&+& 12\tilde{\gamma}A_0^2\frac{\Omega^2}{\gamma} \pi {\rm e}^{-\pi}\sin(2\Omega t_0).
\end{eqnarray}


\begin{thebibliography}{99}

%1
\bibitem{Melnikov1963}  V.K. Melnikov, On the stability of the center for time periodic
perturbations,
Trans. Moscow Math. Soc. 12  (1963) 1--57.

%2
\bibitem{Guckenheimer1983} J. Guckenheimer, P. Holms, Nonlinear
Oscillations,
Dynamical Systems and Bifurcations of Vectorfields, Springer, New York  
1983.

%3
\bibitem{Wiggins1990} S. Wiggins, Introduction to Applied Nonlinear
Dynamical
Systems and Chaos, Spinger, New York 1990.


%4
\bibitem{Tyrkiel2005} E. Tyrkiel, 
On the role of chaotic saddles in generating chaotic dynamics in nonlinear
driven oscillators, Int. J.  Bifurcation and Chaos 15 (2005) 1215--1238.

%5
\bibitem{Szemplinska1995} W. Szempli\'nska-Stupnicka, The analytical
predictive criteria for chaos and escape in nonlinear oscillators: A
survey, Nonlinear Dynamics
7 (1995) 129-147.





%6
\bibitem{Litak2006}
G. Litak, M. Borowiec, Oscillators with asymmetric single and double well
potentials: Transition to chaos revisited, Acta Mechanica 184 (2006) 47-­59.

%7
\bibitem{Litak2007}
G. Litak, A. Syta, M. Borowiec, Suppression of chaos by weak
resonant excitations in a nonlinear oscillator with a non-symmetric
potential, Chaos, Solitons  \& Fractals 
32 (2007) 694--701.

%8
\bibitem{Szemplinska1993} W. Szempli\'nska-Stupnicka, J. Rudowski,
Bifurcations phenomena in a nonlinear oscillator: Approximate analytical
studies versus computer simulation results, Physica D 66 (1993)
368--380.


%9
\bibitem{Kapitaniak1993} T. Kapitaniak, Analytical method of controling 
chaos in Duffing
oscillator, Journal of
Sound and Vibration 163 (1993) 182--187.

%10
\bibitem{Kapitaniak1991} T. Kapitaniak, Chaotic Oscillations in
Mechanical Systems,
Manchester University Press, Manchester 1991.


%\bibitem{Sprott2003} J.C. Sprott, Chaos and Time-Series Analysis,
%Oxford Univ. Press, Oxford 2003.

%11
\bibitem{Ueda1979} Y. Ueda, Randomly transitional phenomena in the system 
governed 
by Duffing's equation, Journal of Statistical Physics 20 (1979) 
181--196.



%12
\bibitem{Chakraborty2006} G. Chakraborty, A conjecture on route to chaos 
in a hard Duffing 
oscillator by homoclinic entanglement, J. Sound Vibr. 294 (2006) 235--440. 

%13
\bibitem{Litak1999}
G. Litak, G. Spuz-Szpos, K. Szabelski, J. Warmi\'nski, Vibration analysis
of self-excited system with parametric forcing and nonlinear stiffness,
Int. J. Bifurcation and Chaos 9 (1999) 493--504.



\end{thebibliography}
\end{document}